\documentclass[aps,pra,paper,shownopacs,superscriptaddress,twocolumn]{revtex4-2}
\usepackage[utf8]{inputenc}
\usepackage{amsmath,amssymb,physics}
\usepackage[hypertexnames=false]{hyperref}
\usepackage{datetime}
\usepackage{enumitem}
\usepackage{graphicx}
\usepackage{braket}
\usepackage{esdiff}
\usepackage{siunitx}
\usepackage{outlines}
\usepackage[normalem]{ulem}
\usepackage[dvipsnames]{xcolor}
\usepackage{xfrac}

\hypersetup{
    colorlinks,
    linkcolor={red!50!black},
    citecolor={blue!50!black},
    urlcolor={blue!80!black}
}

\begin{document}

\title{Deterministic photon waveform adaptation for quantum connectivity}

\author{Jeffrey Mohan}
\affiliation{Welinq, 14 rue Jean Mac\'e, 75011 Paris, France}
\author{J\'er\'emy Berroir}
\affiliation{Welinq, 14 rue Jean Mac\'e, 75011 Paris, France}
\author{Filippo Borselli}
\affiliation{Welinq, 14 rue Jean Mac\'e, 75011 Paris, France}
\author{David Libault}
\affiliation{Welinq, 14 rue Jean Mac\'e, 75011 Paris, France}
\author{Ferhat Loubar}
\affiliation{Welinq, 14 rue Jean Mac\'e, 75011 Paris, France}
\author{Kilian M\"uller}
\affiliation{Welinq, 14 rue Jean Mac\'e, 75011 Paris, France}
\author{Jed Rowland}
\affiliation{Welinq, 14 rue Jean Mac\'e, 75011 Paris, France}
\author{F\'elix Hoffet}
\affiliation{ICFO - Institut de Ciencies Fotoniques, The Barcelona Institute of Science and Technology, Barcelona, Spain.}
\author{Eleni Diamanti}
\affiliation{Welinq, 14 rue Jean Mac\'e, 75011 Paris, France}
\affiliation{LIP6, CNRS, Sorbonne Universit\'e, 75005 Paris, France}
\author{Tom Darras}
\affiliation{Welinq, 14 rue Jean Mac\'e, 75011 Paris, France}
\author{Tommaso Mazzoni}
\email{tommaso.mazzoni@welinq.fr}
\affiliation{Welinq, 14 rue Jean Mac\'e, 75011 Paris, France}
\author{Julien Laurat}
\affiliation{Welinq, 14 rue Jean Mac\'e, 75011 Paris, France}
\affiliation{Laboratoire Kastler Brossel, Sorbonne Universit\'e, CNRS, ENS-Universit\'e PSL, Coll\`ege de France, 4 Place Jussieu, 75005 Paris, France}


\maketitle

\textbf{The scalability of quantum technologies will depend on the ability to interconnect independent quantum systems through photonic channels. However, heterogeneous quantum platforms emit and absorb photons with widely differing properties, severely limiting inter-node interference and modular connectivity. Here we demonstrate a cold-atom optical quantum memory that simultaneously achieves near-unity storage-and-retrieval efficiency and deterministic temporal adaptation of single photons between arbitrary and programmable input and output pulse waveforms. Operating at high optical depth and within a fully integrated architecture, the system can store photons with durations spanning over three orders of magnitude and reshape them arbitrarily without compromising efficiency, achieving compatibility with many current platforms. By augmenting the role of a quantum memory from a passive storage element to an active programmable photonic interface, our results establish a key building block for scalable entanglement-based quantum networks and modular quantum computing architectures.}\\

\begin{figure}[b!]
    \centering
    \includegraphics[width=0.95\columnwidth]{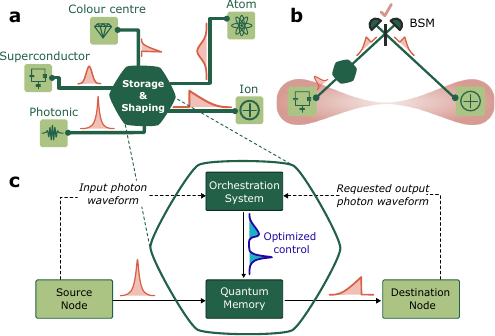}
    \caption{\textbf{Programmable photonic interface for quantum connectivity.} \textbf{a,} Photons emitted by heterogeneous quantum platforms exhibit incompatible temporal waveforms. A programmable quantum memory reshapes them into matching waveforms, enabling photonic interconnects between heterogeneous nodes. \textbf{b,} Schematic of interference-based entanglement generation via Bell-state measurement (BSM) between two nodes, illustrating a key application of photon adaptation. \textbf{c,} For given input and target output waveforms, the temporal profile of the control beam power is computed and applied to optimally store and reshape the photon.}
    \label{fig:fig1}
\end{figure}

\begin{figure*}[t!]
    \centering
    \includegraphics[width=0.8\columnwidth]{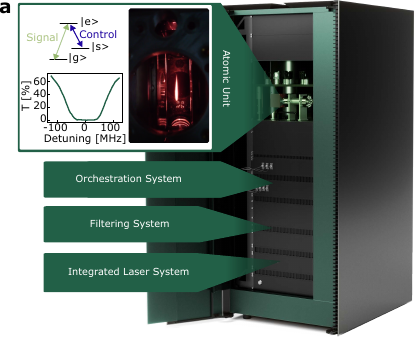}
    \includegraphics[width=\columnwidth]{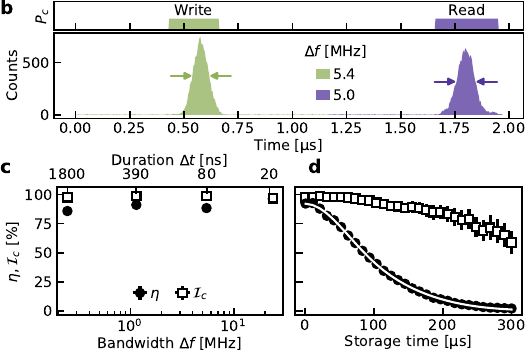}
    \caption{\textbf{QDrive quantum memory system.} \textbf{a,} Schematic of the manufactured integrated system, which we name QDrive. The implementation relies on a large ensemble of \textsuperscript{87}Rb atoms trapped in a vertically-oriented elongated magneto-optical trap. Storage and retrieval are realised via EIT using a control beam with programmable power. The plot shows the signal transmission as a function of the detuning, corresponding to an optical depth of about 750. \textbf{b,} Example memory sequence showing the temporal profile of the control field power $P_c$ together with histograms of photodetection events, revealing the input and retrieved signal waveforms. \textbf{c,} Memory storage-and-retrieval efficiency $\eta$ and conditional overlap $\mathcal{I}_c$ as a function of the input pulse bandwidth. Both quantities remain nearly constant over several orders of magnitude; their averages over this range are $\eta=90.4(1.6)\%$ and $\mathcal{I}_c=97.9(2)\%$. The pulse bandwidth $\Delta f$ and duration $\Delta t \approx 0.44/\Delta f$ are defined as the full widths at half maximum of the Gaussian power in frequency and time as indicated by the arrows in \textbf{b}. \textbf{d,} Lifetime measurement for efficiency $\eta$ and conditional overlap $\mathcal{I}_c$ (Appendix). These data were taken with bright pulses and $\Delta f=\SI{2.5}{MHz}$; all other data were collected with coherent pulses with an average photon number of $\approx1$. No background correction was applied to the histograms for plotting but was taken into account to compute efficiencies. The error bars are obtained from the Poissonian error of the photon counting probabilities as well as power fluctuations during data acquisition.}
    \label{fig:fig2}
\end{figure*}

\begin{figure*}[t!]
    \centering
    \includegraphics[width=1.95\columnwidth]{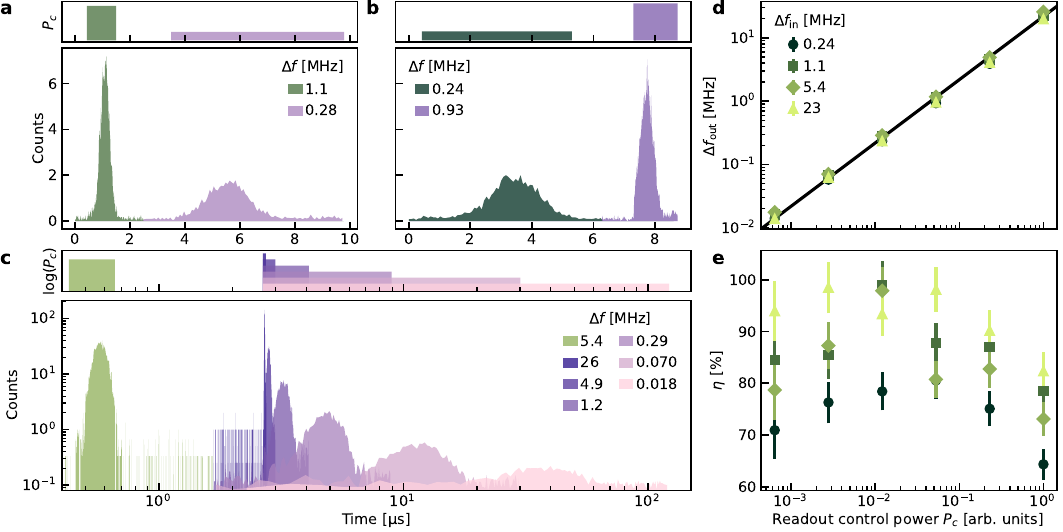}
    \caption{\textbf{Bandwidth adaptation of single-photon Gaussian pulses.} \textbf{a-b,} Examples of pulse stretching and compression. When the read pulse has lower power and longer duration than the write pulse, the signal pulse bandwidth is reduced and the pulse is temporally stretched. Conversely, when the read pulse has higher power and shorter duration than the write pulse, the bandwidth increases and the pulse is compressed. \textbf{c,} By simultaneously varying the power and duration of the read pulse while keeping their product constant, the same input pulse can be either compressed or stretched in time, providing over three decades of tunable output bandwidths. \textbf{d,} Bandwidth of the output pulse $\Delta f_\mathrm{out}$ as a function of the power of the read pulse $P_c$ (along with the appropriate duration) for different input bandwidths. \textbf{e,} The efficiency of the memory is largely independent of the input and output bandwidths over the large explored range, with a slight reduction for the lowest input bandwidth (circles) due to noise in the phase lock between the signal and control lasers. The control power is calibrated such that $P_c=1$ corresponds to $\SI{81(1)}{mW}$ and a Rabi frequency $\Omega/2\pi = \SI{90(1)}{MHz}$ (Appendix). Data were collected with coherent pulses with an average photon number of $\approx1$. The error bars are obtained from the Poissonian error of the photon counting probabilities as well as power fluctuations during data acquisition.}
    \label{fig:fig3}
\end{figure*}

\begin{figure}[t!]
    \centering
    \includegraphics[width=0.95\columnwidth]{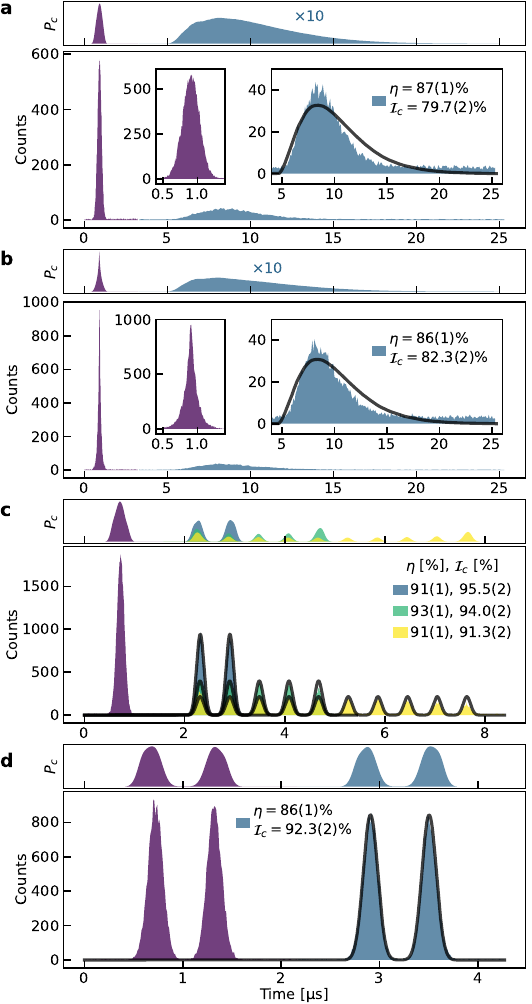}
    \caption{\textbf{High-efficiency storage and generation of arbitrary waveforms.} We compute and implement the temporal profile of the control field power $P_c$ during both the write and read stages to write an arbitrary input and deterministically shape the retrieved signal pulse while maintaining high memory efficiency $\eta$ and conditional overlap between the retrieved and target waveforms $\mathcal{I}_c$. \textbf{a,} A Gaussian input pulse ($\Delta t=\SI{285}{ns}$) is read out as a non-Gaussian waveform ($\Delta t=\SI{4.1}{\micro s}$) characteristic of photons emitted by an ion embedded in a cavity; the grey line in the inset shows the target waveform. \textbf{b,} A double-exponential waveform ($\Delta t=\SI{110}{ns}$) characteristic of single photons generated by cavity-enhanced spontaneous parametric down-conversion is read out as the same target waveform as in \textbf{a}. \textbf{c,} A single Gaussian pulse ($\Delta t=\SI{175}{ns}$) is read out as two, five, or ten Gaussian pulses, demonstrating programmable temporal multiplexing. \textbf{d,} Storage and retrieval of two Gaussian pulses ($\Delta t=\SI{175}{ns}$) for time-bin encoding. Data were collected with coherent pulses with an average photon number of $\approx1$. \vspace{-1mm}}
    \label{fig:fig4}
\end{figure}

While individual quantum platforms have achieved remarkable advances and continue to progress rapidly, intrinsic physical constraints limit the size and performance of monolithic devices. Scalable quantum computing is therefore increasingly envisioned in the form of distributed or data-center architectures~\cite{caleffiDistributedQuantumComputing2024, barralReviewDistributedQuantum2025, shapourianQuantumDataCenter2025}, in which multiple quantum processing units are interconnected through photonic links~\cite{kurizkiQuantumTechnologiesHybrid2015, awschalomDevelopmentQuantumInterconnects2021}. In this framework, photons act as flying carriers that mediate entanglement and coherent interactions between otherwise isolated modules, and distinct qubit modalities can be combined to leverage complementary advantages, for example in speed or gate fidelity. The ability to connect complementary components into unified, reconfigurable structures would enable quantum technologies to move beyond the constraints of individual platforms and lays the foundation for scalable, modular quantum computing~\cite{diamantiQuantumConnectivityScale2026}.

Distinct platforms emit and absorb photons with widely different wavelengths and temporal characteristics as illustrated in Fig.~\ref{fig:fig1}a, presenting a major challenge for their interconnection. For example, cavity-coupled ions emit microseconds-long photons, parametric sources operate in the hundreds-of-nanoseconds regime with double-exponential temporal profiles, and solid-state emitters span even broader bandwidths. While quantum frequency conversion~\cite{kumarQuantumFrequencyConversion1990, tanzilliPhotonicQuantumInformation2005} can address wavelength offsets, correcting temporal differences is also of substantial importance since high-fidelity entanglement swapping, the core primitive underlying distributed architectures, requires indistinguishability of interfering photons in all degrees of freedom~\cite{ManipulatingColorShape2012}. When two platforms emit single photons with temporal mode overlap $\mathcal{J}<1$, the fidelity of interference-based entanglement generation, sketched in Fig.~\ref{fig:fig1}b, is bounded by $(1+\mathcal{J}^2)/2$. Temporal filtering can restore high fidelity, but at the cost of reducing the success rate, which scales as $\mathcal{J}^2$~\cite{craddockQuantumInterference2019, hoffetNearUnityIndistinguishabilitySingle2024, cussenotUnitingQuantumProcessing2025}. In distributed quantum architectures, efficient storage of photonic states is also essential to synchronise processes and allocate resources on demand across the network. Losses and inefficiencies compound across successive operations, severely limiting overall performance. These constraints therefore point to the need for a photonic interface capable of combining high-efficiency storage with deterministic control of photon waveforms, as illustrated in Fig.~\ref{fig:fig1}c.

Atomic ensembles have long been proposed as single-pass platforms for coherent storage and control of photonic wave packets~\cite{hammererQuantumInterfaceLight2010,sangouardQuantumRepeatersBased2011, leiQuantumOpticalMemory2023}. Seminal experiments showed that stored light can, in principle, be retrieved into arbitrary temporal modes~\cite{novikovaOptimalControlLight2007, novikovaOptimalLightStorage2008, phillipsOptimalLightStorage2008}. In practice, however, temporal programmability and high efficiency proved difficult to obtain. Previous demonstrations of pulse shaping operated far below the efficiencies required for scalable architectures~\cite{saglamyurekCoherentStorageManipulation2018, zhangAtomicInterfaceHighDimensional2026}, while high-efficiency memories lacked deterministic and programmable waveform control~\cite{caoEfficientReversibleEntanglement2020, mamannQuantumCryptographyIntegrating2025, wangEfficientQuantumMemory2019}. Other approaches such as absorptive memories with single emitters~\cite{morinDeterministicShapingReshaping2019}, emissive-only memories with atomic ensembles~\cite{eisamanShapingQuantumPulses2004, balicGenerationPairedPhotons2005, duSubnaturalLinewidthBiphotons2008, farreraGenerationSinglePhotons2016}, or electro-optic modulation for biphoton pairs ~\cite{kolchinElectroOpticModulationSingle2008} have also been explored with limited performance or programmability. As a result, the combination of near-lossless storage and full temporal adaptation has remained experimentally out of reach.

In this paper, we overcome these limitations to demonstrate a cold-atom quantum memory that features deterministic and programmable temporal adaptation of single-photon pulses while maintaining near-unity storage-and-retrieval efficiency. The memory operates using electromagnetically induced transparency (EIT) in which the temporal profile of the control beam power $P_c(t)$ governs the mapping between the optical field and the collective spin-wave excitation through dynamic modulation of the pulse group velocity. By engineering $P_c(t)$ with methods from optimal control theory, we leverage complete programmability over the pulse shape during storage and retrieval and demonstrate efficient and deterministic mapping onto arbitrary target temporal waveforms.

Using this mechanism, we efficiently store photon pulses as short as \SI{20}{ns} and generate pulses as long as $\SI{30}{\micro s}$, demonstrating the largest bandwidth achieved with an EIT-based quantum memory. More generally, we store arbitrary waveforms and programmably reshape them during retrieval over a large range of pulse durations and waveforms. This capability enables conversion between photon waveforms that are characteristic of different quantum platforms. As representative examples, we store Gaussian and exponentially peaked pulses with $\sim\SI{1}{MHz}$ bandwidth typical of superconducting~\cite{sekineMicrowavetoOpticalQuantumTransduction2025} and photonic~\cite{feketeUltranarrowBandPhotonPairSource2013} sources and transform them into a non-Gaussian waveform with $\sim\SI{10}{kHz}$ bandwidth that maximises overlap with photons emitted by ion-based systems~\cite{krutyanskiyEntanglementTrappedIonQubits2023}. This work establishes quantum memories as high-efficiency, programmable photonic interfaces capable of adapting photon waveforms across heterogeneous quantum systems, a key functionality for distributed quantum architectures.

These results were made possible by the simultaneous realisation of ultra-high optical depth, low temperature, and effective magnetic-field cancellation, which together allow high-efficiency operation over a wide range of bandwidths and pulse shapes. The robust engineering of the system was essential to achieving this performance, as it enabled precise optimization of the ensemble preparation and access to an extended region of parameter space. The use of optimal control techniques further enabled full exploitation of this regime.\\

\noindent\textbf{An integrated cold-atom quantum memory}\\
\noindent Our implementation is illustrated in Fig.~\ref{fig:fig2}a. It consists of a fully integrated and deployable system enclosed in a transportable, standard 19-inch rack operating at room temperature. The platform hosts an integrated laser system for cooling and trapping, control electronics and an embedded computer, optical benches for light injection and filtering, and a magnetically-shielded ultra-high vacuum chamber housing the cloud of neutral \textsuperscript{87}Rb atoms that serves as the storage medium (Methods and Appendix). Using optimised laser cooling and trapping techniques, we prepare atomic ensembles in an elongated magneto-optical trap with typical optical depths of 750 and temperatures of \SI{30}{\micro K} at a repetition rate of \SI{14}{Hz}. The integrated platform provides fibre-coupled input and output, enabling direct interfacing with photonic quantum network links, and supports autonomous operation over several days (Appendix).

Storage and shaped retrieval are implemented using dynamic EIT~\cite{lukinColloquiumTrappingManipulating2003} on the $D_1$ line at \SI{795}{nm}. The signal (control) field is resonant with the $F=1 \leftrightarrow F'=2$ ($F=2 \leftrightarrow F'=2$) transition. Both Gaussian beams have the same circular polarization and propagate vertically through the atomic cloud with a relative angle of $0.6^\circ$. In this work, the signal and control fields are derived from two external cavity diode lasers and amplitude-modulated with acousto-optic modulators. The signal field is offset-locked to the control and attenuated to produce pulses containing $\approx1$ photon on average.

A typical memory sequence is shown in Fig.~\ref{fig:fig2}b. An incoming Gaussian signal pulse $E_\mathrm{in}(t)$ is written into the memory as a collective spin wave by applying a square pulse of control light with power $P_c$ synchronised to the input signal pulse. After a programmable storage time, the spin wave is read out as an output pulse $E_\mathrm{out}(t)$ using a second control pulse with the same power $P_c$. We characterise the performance of the memory by its storage-and-retrieval efficiency $\eta$ and by the overlap integral $\mathcal{I}_c$ of the output pulse with a target waveform $E_\mathrm{targ}(t)$ ($E_\mathrm{targ} = E_\mathrm{in}$ in this example) conditioned on the detection of a photon (Appendix).

In Fig.~\ref{fig:fig2}c, we benchmark the performance of the system for different signal pulse bandwidths. The storage-and-retrieval efficiency achieves $\eta \gtrsim 90\%$ while the conditional overlap $\mathcal{I}_c$ approaches unity, among the highest values reported for quantum memories. Importantly, both the efficiency and conditional overlap remain nearly constant across a bandwidth range spanning more than three orders of magnitude. The power and duration of the write pulse is tuned to maximise the efficiency for each bandwidth $\Delta f$. This range encompasses the pulse durations typical of many quantum technologies. Figure~\ref{fig:fig2}d shows $\eta$ and $\mathcal{I}_c$ as functions of the storage time. We obtain a lifetime of about $\SI{150}{\micro s}$, which can be extended to the millisecond regime with further cooling of the atoms and suppression of residual magnetic fields (Appendix).\\

\noindent\textbf{Bandwidth adaptation}\\
\noindent In the previous examples, the input and output pulses had the same bandwidth $\Delta f$. We now turn to bandwidth adaptation. Once the signal is mapped onto a spin wave, the excitation no longer retains information about the bandwidth of the input pulse. We exploit this property to change the bandwidth of the output pulse by reading out with a different control power $P_c$ and pulse duration than was used for writing. Figures~\ref{fig:fig3}a and~\ref{fig:fig3}b show two examples where the pulse is either stretched (decreased bandwidth) by reading with lower power and longer duration than in the write pulse, or compressed (increased bandwidth) by reading with higher power and shorter duration. Additional examples are given in Fig.~\ref{fig:fig3}c over orders of magnitude range in the adapted bandwidth. In all cases, the energy of the read pulse, that is the product of its power and duration, is kept equal to that of the write pulse.

Figure~\ref{fig:fig3}d shows that the bandwidth of the retrieved pulse $\Delta f_\mathrm{out}$ depends linearly on the read power $P_c$ independently of the input bandwidth $\Delta f_\mathrm{in}$. This behaviour is due to the spin wave losing the memory of the input pulse bandwidth, and the fact that the bandwidth of the EIT transparency window scales linearly with $P_c$~\cite{lukinColloquiumTrappingManipulating2003}. Moreover, as shown in Fig.~\ref{fig:fig3}e, the efficiency for each input bandwidth $\Delta f_\mathrm{in}$ varies only slightly over the full range of output bandwidths $\Delta f_\mathrm{out}$ explored here. The minimum accessible bandwidth is limited by the memory lifetime, while the maximum bandwidth is limited by the response time of the control beam's amplitude modulator. The input pulse with the smallest bandwidth has a lower efficiency on average due to noise in the offset lock between the signal and control and is not intrinsic to the memory. This noise is irrelevant for higher bandwidth pulses as the associated frequency broadening is much smaller than the wider transparency window bandwidths.\\

\noindent\textbf{Arbitrary waveform storage and generation}\\
\noindent We have shown that the readout power $P_c$ provides direct control over the bandwidth of the retrieved pulse while maintaining high efficiency. Temporal modulation of the control power during writing and reading therefore enables arbitrary waveform storage and generation via its direct influence on the group velocity~\cite{gorshkovPhotonStorageLambdatype2007a, gorshkovPhotonStorageLambdatype2008}. By modulating the control field during the write stage, the memory can efficiently absorb non-Gaussian input pulses, while modulation during retrieval enables deterministic shaping of the output waveform.

To implement the storage and shaping of arbitrary photon waveforms, we developed an algorithm that computes \textit{ab initio} the optimal temporal profile of the control field power $P_c(t)$ during both the write and read stages (Methods and Appendix). We compute the profiles during both stages simultaneously by numerically optimising the unconditional overlap integral $\mathcal{I} = \eta \mathcal{I}_c$ using techniques from optimal control theory. By solving the Maxwell-Bloch equations governing the coupled atomic and photonic dynamics with a trial $P_c(t)$ and iteratively updating it via gradient ascent, the algorithm rapidly converges to an optimal profile that simultaneously maximises both $\eta$ and $\mathcal{I}_c$. In contrast to earlier approaches~\cite{novikovaOptimalLightStorage2008, gorshkovPhotonStorageLambdatype2007a, phillipsOptimalLightStorage2008} which optimised the write and read processes sequentially and relied on intermediate spin-wave optimisation, this method streamlines the process by directly determining the control fields that maximise the overall mapping between input and target temporal modes. Additional details of the optimisation procedure, together with different examples of representative spin-wave dynamics, are provided in the Appendix.

We demonstrate the performance of the protocol with several examples shown in Fig.~\ref{fig:fig4}. The optimised temporal profile of the control field power $P_c(t)$ used during the write and read stages is displayed in the upper panel for each case. Figure~\ref{fig:fig4}a shows the storage of a Gaussian input pulse and its retrieval as a stretched, non-Gaussian output pulse. The target waveform corresponds to that of a photon emitted by a single calcium ion embedded in a cavity, as reported in~\cite{krutyanskiyEntanglementTrappedIonQubits2023, cussenotUnitingQuantumProcessing2025}, and differs in time scale by over one order of magnitude from the input pulse. Figure~\ref{fig:fig4}b shows the same target but obtained from a non-Gaussian, double-exponential input pulse characteristic of cavity-enhanced spontaneous parametric down conversion~\cite{feketeUltranarrowBandPhotonPairSource2013, tsaiQuantumStorageManipulation2020}. These results demonstrate that the desired output temporal mode can be generated independently of the input waveform while maintaining high efficiency $\eta$ and conditional overlap $\mathcal{I}_c$. Without any reshaping, the input pulse has an overlap of only $\mathcal{I}_c \approx 9\%$ with the target mode in both cases. This waveform adaptation would translate into a 60-fold increase in the entanglement generation rate via Bell-state measurement compared to the absence of reshaping.

We further exploit the programmable control of the readout process to generate multiple temporal modes from a single stored excitation. Figure~\ref{fig:fig4}c shows the conversion of a single Gaussian input pulse into two, five, or ten Gaussian output pulses of equal amplitude and duration. Within the explored range, an arbitrary number of time bins can be generated without significant degradation of $\eta$ or $\mathcal{I}_c$. This capability naturally fits within the broader framework of high-dimensional photonic encoding, in which quantum information is distributed across multiple temporal modes~\cite{singhPhotonicQuantumInformation2025}.

Finally, we demonstrate the storage and retrieval of two successive Gaussian pulses, which can be used for time-bin encoding, as shown in Fig.~\ref{fig:fig4}d. The programmable control of the write and read stages provides a versatile tool for manipulating photonic states encoded in temporal modes. In addition to amplitude shaping as demonstrated here, we anticipate that the phase of the control field can be tuned through the phase of the radio-frequency drive applied to the acousto-optical modulator, enabling phase shifts between temporal modes of the retrieved signal~\cite{morinDeterministicShapingReshaping2019}.

Further improvements in our calculation and implementation of the optimal control waveforms are readily accessible. First, our model does not include any dephasing of the spin wave that leads to the finite memory lifetime, for example thermal motion, magnetic field gradients, and off-resonant photon scattering. Second, the finite response time and limited number of samples of the used pulse generator driving the acousto-optical modulators distort the control waveforms away from the ideal calculated profiles. Third, the photodiode used to calibrate and linearise the modulators of the signal and control was found to exhibit a small nonlinearity, further distorting the waveforms. Even after accounting for these imperfections, the control waveforms can be further refined by iteratively feeding back the experimentally measured output signal to update the control profile~\cite{novikovaOptimalControlLight2007, phillipsOptimalLightStorage2008, guoNearPerfectBroadbandQuantum2025}. With these improvements, simulations indicate that efficiencies exceeding $\eta>94\%$ and conditional overlaps $\mathcal{I}_c>99\%$ can be consistently achieved (Appendix). \\

\noindent\textbf{Outlook}\\
\noindent We have demonstrated a cold-atom quantum memory capable of deterministic temporal adaptation of single-photon pulses while maintaining high storage-and-retrieval efficiency. By engineering the temporal profile of the control beam power, the system achieves programmable photonic waveform reshaping across more than three orders of magnitude in bandwidth with high conditional mode overlap. We further show high-efficiency storage and retrieval of arbitrary temporal waveforms, encompassing temporal mode conversion between disparate photonic sources, temporal multiplexing into high-dimensional time-bin states, and storage of multiple modes simultaneously for time-bin encoding.

Implemented in a fully integrated and deployable platform, this photonic interface addresses a central challenge in distributed quantum architectures: connecting heterogeneous quantum nodes through photonic channels that require precise temporal mode matching. By allowing a single system to dynamically adapt to the emission characteristics of diverse quantum emitters, such interfaces provide a practical route towards coherent interconnects between otherwise incompatible quantum platforms. More broadly, programmable quantum memories of this kind could enable flexible photonic interconnect layers for emerging quantum networks, supporting entanglement distribution, modular quantum computing, and high-dimensional photonic information processing. As quantum technologies continue to diversify, such adaptive interfaces may play a central role in linking quantum systems into scalable distributed architectures.

\vspace{0.8cm}
\noindent{\fontfamily{phv}\selectfont{\color{black}\textbf{METHODS}}\\}

\noindent \textbf{Atomic ensemble preparation} \\
Each cycle begins with \SI{60}{ms} of loading the magneto-optical trap (MOT) from rubidium dispensers. The quadrupole coils are designed to produce a stronger gradient along the two transverse directions than along the vertical direction, the optical axis of the signal and control beams, to achieve a cigar-shaped centimetre-long cloud that maximises the optical depth. We then implement a dark, compressed MOT for \SI{10}{ms} to increase the density by reducing the power of the repumper and ramping up the magnetic field. We then rapidly switch off the magnetic field and cool for \SI{3}{ms} with a $\Lambda$-enhanced grey molasses on the $D_2$ line~\cite{rosiLambdaenhancedGreyMolasses2018} and optically pump all atoms into the $\ket{5S_{1/2}, F=1}$ level. At the end of these \SI{73}{ms} of preparation, the system typically has an OD of 750 and a temperature of \SI{30}{\micro K}. At this stage, the system is ready to perform memory sequences for the thermally-limited lifetime of the OD, typically \SI{10}{ms}. In addition to magnetically shielding the vacuum system, we use three pairs of Helmholtz coils inside the shield to zero the residual magnetic field induced by the eddy currents that are produced by the fast switch-off of the quadrupole field.\\

\noindent \textbf{EIT laser system and pulse shaping} \\
The control and signal beams are derived from two diode lasers (Toptica, DL Pro). The control is locked to a rubidium vapour cell and amplified with a tapered amplifier (Toptica, BoosTA), while the signal is offset-locked to the control (Vescent, D2-135). The signal and control beam waists on the atoms are \SI{120}{\micro m} and \SI{620}{\micro m} respectively. We use acousto-optic modulators (AOMs) (AA Opto-Electronic) operated at \SI{250}{MHz} and \SI{204}{MHz} for the signal and control respectively with 10\%-90\% response times of \SI{7}{ns} for switching and power modulation. We linearised the AOMs and their driving electronics (M-Labs, Urukul; Mini-Circuits, ZASWA2-50DR-FA+; AA Opto-Electronic, AMPB-B-34-10.500) by measuring the optical power as a function of the amplitude of their radio-frequency drives and inverting the resulting lookup table. The avalanche photodiode (Thorlabs, APD431A) used to measure the power exhibits a residual nonlinearity, which slightly degrades this linearisation. We do not correct for the finite response time of the AOMs.\\

\noindent \textbf{Detection system} \\
Before detecting the retrieved photons, we direct the memory output to two custom Fabry-Perot etalons in series to filter the control light out from the signal (see Appendix). We then detect the signal photons transmitted through the cavities using a single-photon counting module (Excelitas, SPCM-800-44-FC) connected to a time tagger (Swabian Instruments, Time Tagger 20). For the longest duration histograms, the recorded histograms have been re-binned by consolidating a few adjacent bins to improve visualisation. The signals would otherwise be buried by the dark count rate of the single-photon counting module $\approx \SI{100}{s^{-1}}$. To accurately measure the storage-and-retrieval efficiency, we first collect a reference pulse by running the memory sequence with only signal light and no atoms to measure the input pulse shape. We then repeat the sequence with only control light with atoms to determine the amount of control light leakage. Finally, we run the full memory sequence with signal, control, and atoms. We repeat this whole three-step sequence with no atoms to characterise and correct for any systematic bias in the  pulse energies in any of these stages.\\

\noindent \textbf{Optimising the control waveforms} \\
Given the waveform of the input signal $E_\mathrm{in}(t)$ and the waveform of the control Rabi frequency $\Omega(t)$ over the write, storage, and readout stages, we can solve the Maxwell-Bloch equations of motion for the coupled dynamics of the electric field, spin wave, and atomic polarisation to find the electric field at the output of the atomic ensemble $E_\mathrm{out}(t)$. Our objective is to generate an arbitrary and programmable target output waveform $E_\mathrm{out}(t) = E_\mathrm{targ}(t)$. To achieve this, we optimise the control field $\Omega(t)$ to maximise the unconditional overlap integral $\mathcal{I} = \eta \mathcal{I}_c$. Specifically, we transform the optimisation of $\mathcal{I}$ under the constraints of the equations of motion into an unconstrained optimisation of a new figure of merit proportional to $\mathcal{I}$ by introducing a Lagrange multiplier for each equation of motion. The gradient of this new figure of merit with respect to $\Omega(t)$, which can be computed by solving the original and conjugate equations of motion, can then be used to improve $\Omega(t)$. The algorithm therefore proceeds by making an initial guess for $\Omega(t)$, solving the equations of motion, updating $\Omega(t)$ via gradient ascent, and repeating the process until convergence of the figure of merit. The complete mathematical framework is detailed in the Appendix.

\clearpage
\bibliographystyle{naturemag}

\vspace{0.5cm}
\noindent \textbf{Acknowledgments} We thank Tridib Ray, Mohamed Guessoum, Isabelle Riou, and Nahel Bentaj for their contributions at an early stage of the project. This work was supported by the European Union's Horizon Europe research and innovation programme via the Flagship project QIA (No. 101102140 and No. 101297140) and the EIC Accelerator project No. 101188682, and by the French National Research Agency via the France 2030 project QMemo (ANR-22-PETQ-0010). J.L. is a member of the Institut Universitaire de France.\\

\noindent \textbf{Author contributions} J.M., J.B., F.B., D.L., F.L., K.M., and J.R. contributed to the development of the system, including the integration of all the hardware components, under the close supervision of T.M.; J.B., J.M., and T.M. developed the orchestration stack; J.M. and J.B. developed the optimisation algorithm; J.M. performed the data taking and analysis supervised by T.M; F.H. contributed to early discussions on mode adaptation; F.L. designed the filtering cavities; T.M., E.D., T.D., and J.L. supervised the overall implementation; J.M., J.B., T.M., and J.L wrote the paper with input from all co-authors.\\

\clearpage
\renewcommand{\thefigure}{A\arabic{figure}}
\renewcommand{\thetable}{A\arabic{table}}
\setcounter{figure}{0}

\onecolumngrid

\appendix
\section{System details}

\subsection{Hardware}

\begin{figure}[h]
    \centering
    \includegraphics[height=6cm]{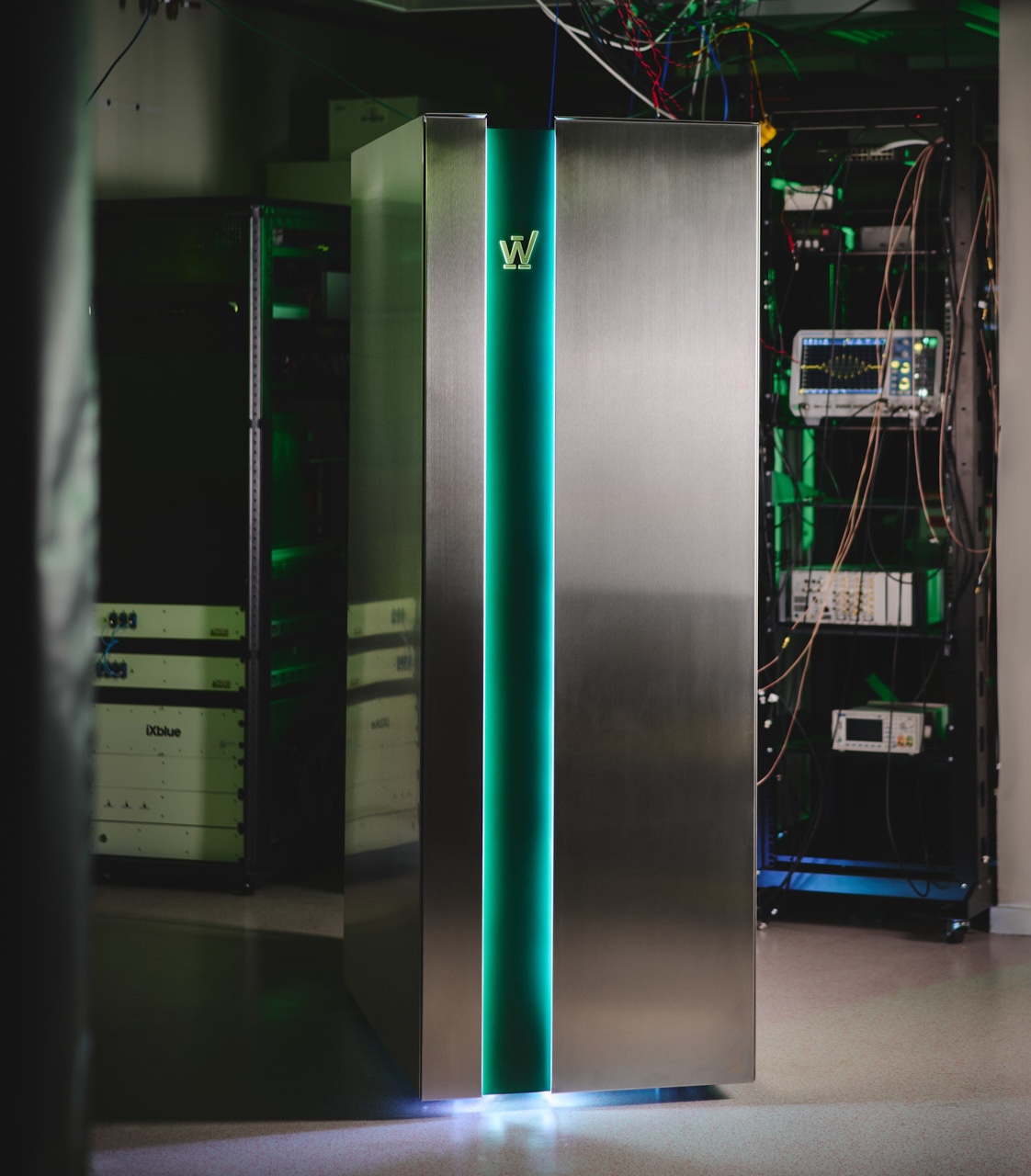}
    \includegraphics[height=6cm]{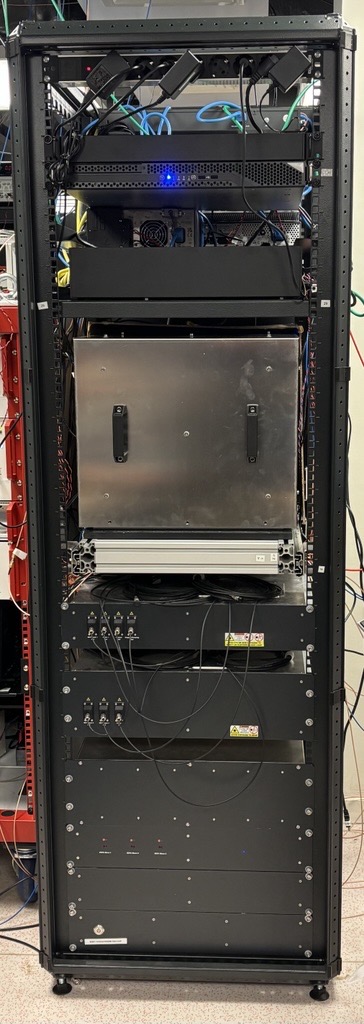}
    \includegraphics[height=5cm]{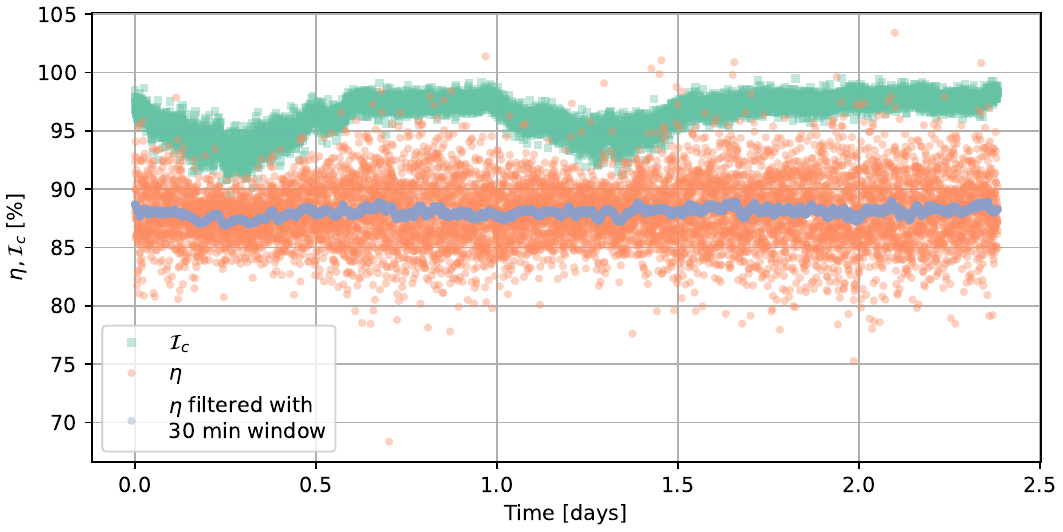}
    \caption{\textbf{QDrive system and its operational stability.} Views of the system with and without its protective enclosure, showing the internal architecture including the integrated laser system, magnetically shielded vacuum system, filter cavities, racked computer running the orchestration system, and control electronics. We measured the storage-and-retrieval efficiency $\eta$ and conditional overlap integral $\mathcal{I}_c$ using bright pulses over several days without interruption or re-calibration. The fluctuations in the passively stable powers of the signal and control beams give rise to the shot-to-shot fluctuations in the efficiency and the long-term drift in $\mathcal{I}_c$.}
    \label{fig:qdrive}
\end{figure}

Figure~\ref{fig:qdrive} displays photos of the QDrive system. At the bottom of the rack is a fully integrated Intelligent Laser System (Exail), which provides over \SI{1}{W} of light at \SI{780}{nm} from power-stabilised fibred outputs for preparing the atomic ensemble. This system is nearly fully fibred, based on distributed-feedback lasers at \SI{1560}{nm} which are amplified by erbium-doped fibre amplifiers, frequency doubled, and power switched and modulated by micro-optic free-space modules. The two cooling lasers and one repumper laser are all are independently phase-locked to a master laser locked to a rubidium spectroscopy cell. The entire physics package is integrated within a small volume of approximately \SI{100}{L}, comprising the vacuum chamber surrounded by MOT coils and collimators, all enclosed within a compact magnetic shield. The free-space EIT beam preparation and collection benches are also included, with the full system mounted on a 19-inch drawer to facilitate easy maintenance. The core of the system's control electronics is based on the Sinara device family (M-Labs). Our orchestration system is built on top of the ARTIQ framework~\cite{bourdeauducqARTIQ2021}. The output of the memory is coupled into a polarisation-maintaining fibre and directed into two custom Fabry-Perot etalons in series to filter the control light out from the signal. The filters feature a combined transmission of 94\% at the signal frequency, a full width at half maximum transmission bandwidth of \SI{300}{MHz}, and \SI{50}{dB} suppression of the control frequency relative to the signal. The transmitted photons are detected using a single-photon counting module (Excelitas, SPCM-800-44-FC) connected to a time tagger (Swabian Instruments, Time Tagger 20). The signal-to-noise ratio, defined by the ratio of signal counts to noise counts, is $\sim10$ and is currently limited by pollution from the control beam at the control frequency as well as at the signal frequency originating from broadband amplified spontaneous emission; this can be improved by additional spectral filtering without significant loss.

\begin{figure}[h]
    \centering
    \begin{minipage}{10cm}
        \centering
        \includegraphics{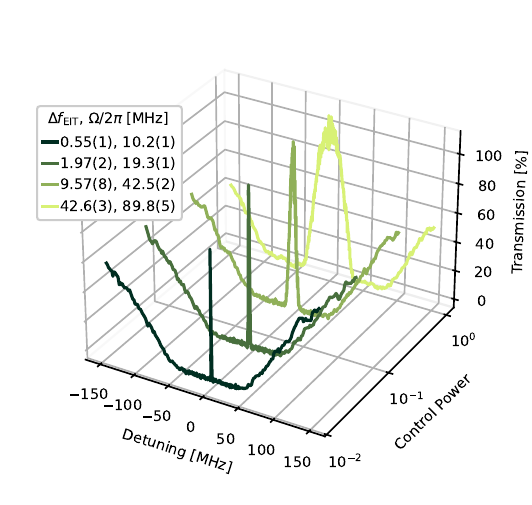}
    \end{minipage}
    \begin{minipage}{7cm}
        \caption{\textbf{EIT spectra with the control powers used for each of the input pulse bandwidths used.} The transparency window bandwidth $\Delta f_\mathrm{EIT} = 2\sqrt{2\log2}\sigma_\mathrm{EIT}$ is given by the width of the Gaussian fit to the window $\sigma_\mathrm{EIT}$, and the Rabi frequency $\Omega = \sqrt{2\pi\sqrt{8\mathrm{OD}}\Gamma\sigma_\mathrm{EIT}}$ is given by the same width, the independently measured OD, and the natural linewidth of the transition $\Gamma/2\pi=\SI{5.75}{MHz}$~\cite{fleischhauerElectromagneticallyInducedTransparency2005}. The reduction in the transmission peak height for the lowest control power is due to the noise in the signal frequency lock. The broadening of the absorption wings for the largest $\Omega$ is due to Autler-Townes splitting of the excited state.}
        \label{fig:eit}
    \end{minipage}
\end{figure}

As demonstrated in Fig.~3 of the main text, the bandwidth of the retrieved pulse is directly linked to the power of the readout control power $P_c$. This is due to the fact that $P_c$ also directly controls the width of the EIT transparency window. We demonstrate this in Sup. Fig.~\ref{fig:eit}, which shows the measured EIT spectra at the optimal control power for each input pulse bandwidth used in this work.

\subsection{Duration and bandwidth of a Gaussian pulse}

We define the electric field of the signal pulse to be $E(t) \propto e^{-(t/2\sigma_t)^2}$ such that the power $P(t) \propto \abs*{E(t)}^2 \propto e^{-t^2/2\sigma_t^2}$ has a characteristic width of $\sigma_t$ and a full-width at half max (FWHM) of $\Delta t = 2\sqrt{2\log2}\sigma_t$. The frequency spectrum of the field is given by its Fourier transform $\tilde{E}(t) = \int E(t) e^{-i2\pi ft} \dd{t} \propto e^{-(2\pi\sigma_t f)^2}$, so the characteristic width of the power spectrum $\tilde{P}(f) \propto \abs*{\tilde{E}(f)}^2 \propto e^{-t^2/2\sigma_f^2}$ is $\sigma_f = 1/4\pi\sigma_t$, and its FWHM is $\Delta f = 2\sqrt{2\log2}\sigma_f = 2\log2/\pi\Delta t \approx 0.44/\Delta t$. We extract these values from our data by fitting a Gaussian to the measured histogram.

\subsection{Memory lifetime}

The efficiency as a function of the storage duration is given in Fig.~2d of the main text. The displayed fit is given by $\eta(t) = \eta_0 \bqty*{x e^{-(t/\tau_1)^2} + (1-x) e^{-(t/\tau_2)^2}}$ with $\eta_0=\SI{92.0(4)}{\%}, x=0.57(3), \tau_1=\SI{154(3)}{\micro s}, \tau_2=\SI{78(3)}{\micro s}$. The two distinct lifetimes reflect dephasing due to thermal atomic motion~\cite{zhaoMillisecondQuantumMemory2009}, which affects both magnetically sensitive ($m_F=\pm1,\pm2$) and insensitive ($m_F=0$) states, as well as dephasing due to inhomogeneous residual magnetic fields, which affect only the magnetically sensitive states. The memory lifetime is currently limited by the magnetic field gradients of the eddy currents, for which we are currently developing solutions. Once these gradients are canceled, the memory lifetime is limited by finite temperature. We have observed by allowing the gradients to decay completely before initiating the memory sequence that the lifetime can be extended to $\tau > \SI{800}{\micro s}$, which can be increased even further by reducing the angle between the probe and control beams.

\section{Computing the control waveform for programmable pulse shaping}

In this section, we describe the algorithm to compute the temporal profiles of the control power $P_c(t)$ used to write and read the desired signal waveforms shown in Fig.~4 of the main text. The procedure builds upon the framework introduced in~\cite{gorshkovPhotonStorageLambdatype2007a, gorshkovPhotonStorageLambdatype2008, novikovaOptimalLightStorage2008}, while extending it through a unified optimal-control formulation that simultaneously optimises the write and read stages.

We consider an ensemble of three-level atoms with a longitudinal density distribution $n(z)$ illuminated by a time-varying input signal field $E_\mathrm{in}(t)$ and spatially uniform control field with Rabi frequency $\Omega(t)$ with single-photon detuning $\Delta$ and two-photon detuning $\delta$. Assuming the amplitude of the signal varies slowly in space and time relative to its vacuum wavelength and frequency, then the electric field $E$, spin-wave $S$, and polarisation $P$ are described by the following classical equations of motion (EoM) and boundary conditions
\begin{align} \label{eq:eom}
    \left\{\begin{array}{ll}
    \partial_{\tilde{z}} E(\tilde{z}, \tilde{t}) &= i\sqrt{\mathrm{OD}/2} P(\tilde{z}, \tilde{t}) \\
    \partial_{\tilde{t}} P(\tilde{z}, \tilde{t}) &= i\sqrt{\mathrm{OD}/2} E(\tilde{z}, \tilde{t}) + i\tilde{\Omega}(\tilde{t}) S(\tilde{z}, \tilde{t}) - (1 + i\tilde{\Delta})P(\tilde{z}, \tilde{t}) \\
    \partial_{\tilde{t}} S(\tilde{z}, \tilde{t}) &= i\tilde{\Omega}^*(\tilde{t}) P(\tilde{z}, \tilde{t}) + i\tilde{\delta} S(\tilde{z}, \tilde{t}) \\
    E(0, \tilde{t}) &= E_\mathrm{in}(\tilde{t}) \\
    P(\tilde{z}, 0) &= 0 \\
    S(\tilde{z}, 0) &= 0.
    \end{array}\right.
\end{align}
The dimensionless variables are the longitudinal coordinate $\tilde{z} = \int_{-\infty}^z n(z) \dd{z}/\int_{-\infty}^\infty n(z) \dd{z}$, time $\tilde{t} = \gamma(t-z/c)$, Rabi frequency $\tilde{\Omega} = \Omega/2\gamma$, single-photon detuning $\tilde{\Delta} = \Delta/\gamma$, and two-photon detuning $\tilde{\delta} = \delta/\gamma$, where $c$ is the speed of light in vacuum and $\gamma/\pi=\SI{5.75}{MHz}$ is the decoherence rate of the excited state (half the natural linewidth of its transitions to the ground states).

The output signal $E_\mathrm{out}(t) = E(1,\tilde{t})$ can be computed from these EoM given a control waveform $\Omega(t)$ using standard methods for numerical integration of partial differential equations. However, we are interested in computing $\Omega(t)$ to achieve a particular $E_\mathrm{out}(t) = E_\mathrm{targ}(t)$ given a particular $E_\mathrm{in}(t)$. We benchmark this retrieval with specific figures of merit, namely the efficiency $\eta$, the overlap integral $\mathcal{I}$, and the conditional overlap integral $\mathcal{I}_c$ defined as
\begin{equation}\label{eq:figures_of_merit}
    \eta = \frac{\int\abs{E_\mathrm{out}(t)}^2 \dd{t}}{\int\abs{E_\mathrm{in}(t)}^2 \dd{t}}, \qquad
    \mathcal{I} = \frac{\abs{\int E_\mathrm{out}(t) E_\mathrm{targ}^*(t) \dd{t}}^2}{\bqty{\int\abs{E_\mathrm{targ}(t)}^2 \dd{t}}^2}, \qq{and}
    \mathcal{I}_c = \mathcal{I}/\eta = \frac{\abs{\int E_\mathrm{out}(t) E_\mathrm{targ}^*(t) \dd{t}}^2}{\int\abs{E_\mathrm{out}(t)}^2 \dd{t} \int\abs{E_\mathrm{targ}(t)}^2 \dd{t}}.
\end{equation}
More precisely, we want to maximise all of these figures of merit under the constraints of the EoM Eq.~\ref{eq:eom}. This problem can be solved by introducing the Lagrange multipliers $\bar{E}(z, t)$, $\bar{P}(z, t)$, and $\bar{S}(z, t)$ to enforce the EoM as constraints on the following functional
\begin{equation}
    \bar{\mathcal{I}} = \mathcal{I} + \int \bar{E}^*(-\partial_{\tilde{z}}E + i\sqrt{\mathrm{OD}/2}P) + \bar{P}^*[-\partial_{\tilde{t}}P + i\sqrt{\mathrm{OD}/2}E + i\tilde{\Omega} S - (1+i\tilde{\Delta})P] + \bar{S}^*(-\partial_{\tilde{t}}S + i\tilde{\Omega} P + i\tilde{\delta}S) + \mathrm{c.c.} \dd{\tilde{z}}\dd{\tilde{t}}.
\end{equation}
The constrained optimisation of $\mathcal{I}$ with respect to $\Omega(\tilde{t})$ is therefore transformed into an unconstrained optimisation of $\bar{\mathcal{I}}$ with respect to $\Omega(\tilde{t})$ where the Lagrange multipliers satisfy the following EoM and boundary conditions
\begin{align} \label{eq:lagrange_eom}
    \left\{\begin{array}{ll}
    \partial_{\tilde{z}} \bar{E}(\tilde{z}, \tilde{t}) &= i\sqrt{\mathrm{OD}/2} \bar{P}(\tilde{z}, \tilde{t}) \\
    \partial_{\tilde{t}} \bar{P}(\tilde{z}, \tilde{t}) &= i\sqrt{\mathrm{OD}/2} \bar{E}(\tilde{z}, \tilde{t}) + i\tilde{\Omega}(\tilde{t}) \bar{S}(\tilde{z}, \tilde{t}) - (1 - i\tilde{\Delta})\bar{P}(\tilde{z}, \tilde{t}) \\
    \partial_{\tilde{t}} \bar{S}(\tilde{z}, \tilde{t}) &= i\tilde{\Omega}^*(\tilde{t}) \bar{P}(\tilde{z}, \tilde{t}) + i\tilde{\delta} \bar{S}(\tilde{z}, \tilde{t}) \\
    \bar{E}(1, \tilde{t}) &= E_\mathrm{targ}(\tilde{t}) \\
    \bar{P}(\tilde{z}, \tilde{t}_\mathrm{max}) &= 0 \\
    \bar{S}(\tilde{z}, \tilde{t}_\mathrm{max}) &= 0.
    \end{array}\right.
\end{align}
Equation~\ref{eq:lagrange_eom} is found by setting the variation of $\bar{\mathcal{I}}$ to 0 for any variation in $E$, $S$, or $P$.

We begin the optimisation procedure with an initial guess for $\Omega(t)$, for which we take low-pass-filtered versions of $E_\mathrm{in}(t)$ during the write stage and $E_\mathrm{targ}(t)$ during the read stage scaled to appropriate peak amplitudes $\tilde{\Omega}_i = 2\sqrt{\mathrm{OD}/\gamma t_i}$ where $t_i$ is the duration of the $i=\mathrm{write/read}$ stage. We then solve the EoM for the fields (Eq.~\ref{eq:eom}) and Lagrange multipliers (Eq.~\ref{eq:lagrange_eom}), compute the figures of merit, $\mathcal{I}$, $\eta$, and $\mathcal{I}_c$, and update the control waveform according to the following gradient ascent rule
\begin{equation} \label{eq:update}
    \tilde{\Omega}(\tilde{t}) \to \tilde{\Omega}(\tilde{t}) + \lambda \pdv{\bar{\mathcal{I}}}{\tilde{\Omega}(\tilde{t})}
\end{equation}
where $\lambda \sim 5$ and the functional derivative of the constrained figure of merit is given by
\begin{equation}
    \pdv{\bar{\mathcal{I}}}{\tilde{\Omega}(\tilde{t})} = -2\int \mathrm{Im}\bqty{\bar{S}^*(\tilde{z}, \tilde{t}) P(\tilde{z}, \tilde{t}) + S(\tilde{z}, \tilde{t}) \bar{P}^*(\tilde{z}, \tilde{t})} \dd{\tilde{z}}.
\end{equation}
We then iterate this procedure of solving the EoM and updating the control waveform for 10--100 iterations until convergence of the figures of merit. To improve convergence and numerical stability, we enforce $\Omega = E_\mathrm{in} = E_\mathrm{targ} = 0$ during the whole storage stage as well as at the first and last time step of both the write and read stages by subtracting a linear gradient from each field in each stage. The optimisation typically takes $\sim\SI{1}{min}$ on a laptop.

We experimentally implement the computed optimal control waveform $\Omega(t)$ by re-sampling the power $P_c(t) \propto \abs{\Omega(t)}^2$ to the sampling grid of the waveform generator driving the AOM of the control beam. We then multiply $P_c(t)$ by a constant fraction of the maximum available control beam power and convert the waveform of the optical power into a waveform of the amplitude of the radio-frequency drive using a calibrated lookup table. We tune the constant multiplication factor to maximise the measured $\mathcal{I}$ with bright pulses. We find that the power that maximises $\mathcal{I}$ does not coincide with that which maximises $\eta$ or $\mathcal{I}_c$, which is likely due to the known issues with the current state of the computation and implementation and not a fundamental feature. We found that allowing for different power scales during the write and read stage does not improve performance. We apply the same linearisation procedure to generate the waveforms for producing the signal pulses.

This model of the system (Eq.~\ref{eq:eom}) neglects all dephasing mechanisms and therefore does not produce the true optimal control waveform. Such effects, for example the thermal motion and magnetic field gradients that give rise to the finite memory lifetime shown in Fig.~2 of the main text, can be incorporated into Eq.~\ref{eq:eom} and Eq.~\ref{eq:lagrange_eom} by adding additional terms to the right hand side of the EoM for the spin wave. This is often done phenomenologically with a term $-\gamma_s S$~\cite{gorshkovPhotonStorageLambdatype2007a}, though such a term leads to exponential decay of the efficiency rather than the Gaussian decay we observe. Dephasing due to a magnetic field gradient on a Gaussian cloud corresponds to a gradient in the two-photon detuning of the magnetically sensitive states $\delta(z) \propto \gamma_s z \propto \gamma_s \mathrm{erfc}^{-1}[2(1-\tilde{z})]$. Photon scattering can be similarly incorporated as it corresponds to an imaginary component of $\delta$~\cite{lukinColloquiumTrappingManipulating2003}. In principle, it is possible to include the effect of dephasing due to thermal motion, though it is less straightforward as the model does not include the motional degrees of freedom of the atoms whose fluctuations are responsible for this dephasing.

\begin{figure}[h]
    \centering
    \includegraphics[width=\textwidth]{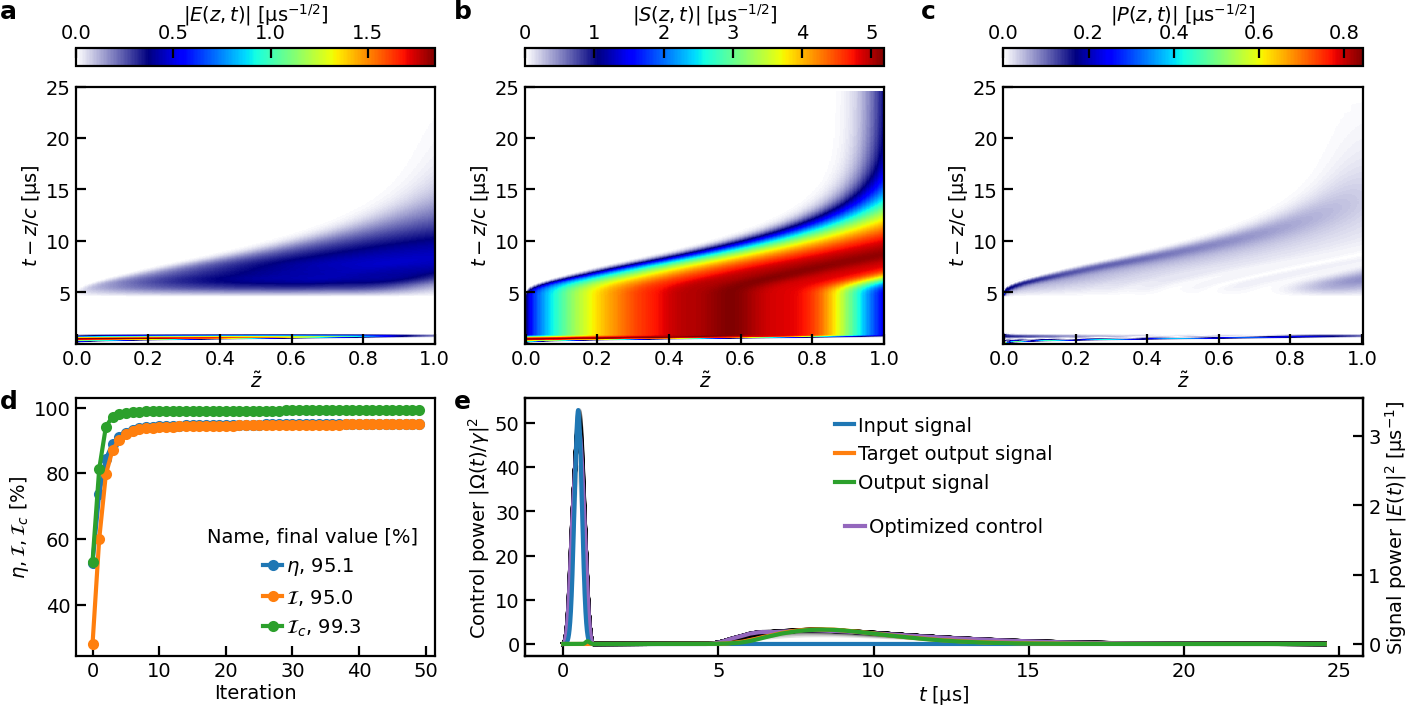}
    \caption{\textbf{Computing the optimal control waveforms and simulating system dynamics for the arbitrary pulse storage and generation in Fig.~4a of the main text.} \textbf{a-c,} The electric, spin-wave, and polarisation fields with the optimised control waveform over the full memory sequence. \textbf{d,} The figures of merit at each iteration of the optimisation procedure. \textbf{e,} The input, target, and output signal waveforms. The control waveform at each iteration is shown in grey, with lighter colours corresponding to earlier iterations, and the final iteration in purple. The evolution of the control profile is more clearly visible in Sup. Fig.~\ref{fig:timebin}e.}
    \label{fig:gauss_to_ion}
\end{figure}

\begin{figure}[h]
    \centering
    \includegraphics[width=\textwidth]{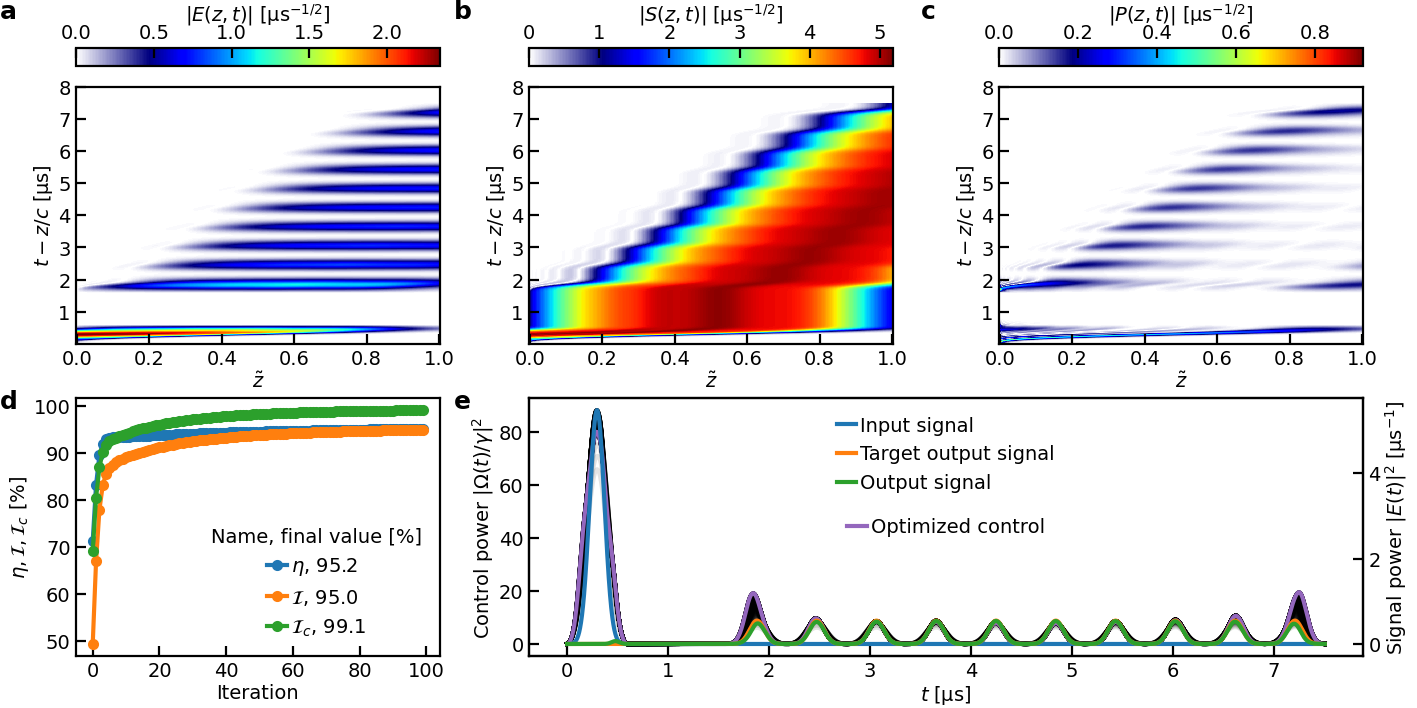}
    \caption{\textbf{Optimal waveforms and system dynamics corresponding to the ten output pulses in Fig.~4c of the main text.} See the caption of Sup. Fig.~\ref{fig:gauss_to_ion} for details.}
    \label{fig:gauss_to_10_gauss}
\end{figure}

\begin{figure}[h]
    \centering
    \includegraphics[width=\textwidth]{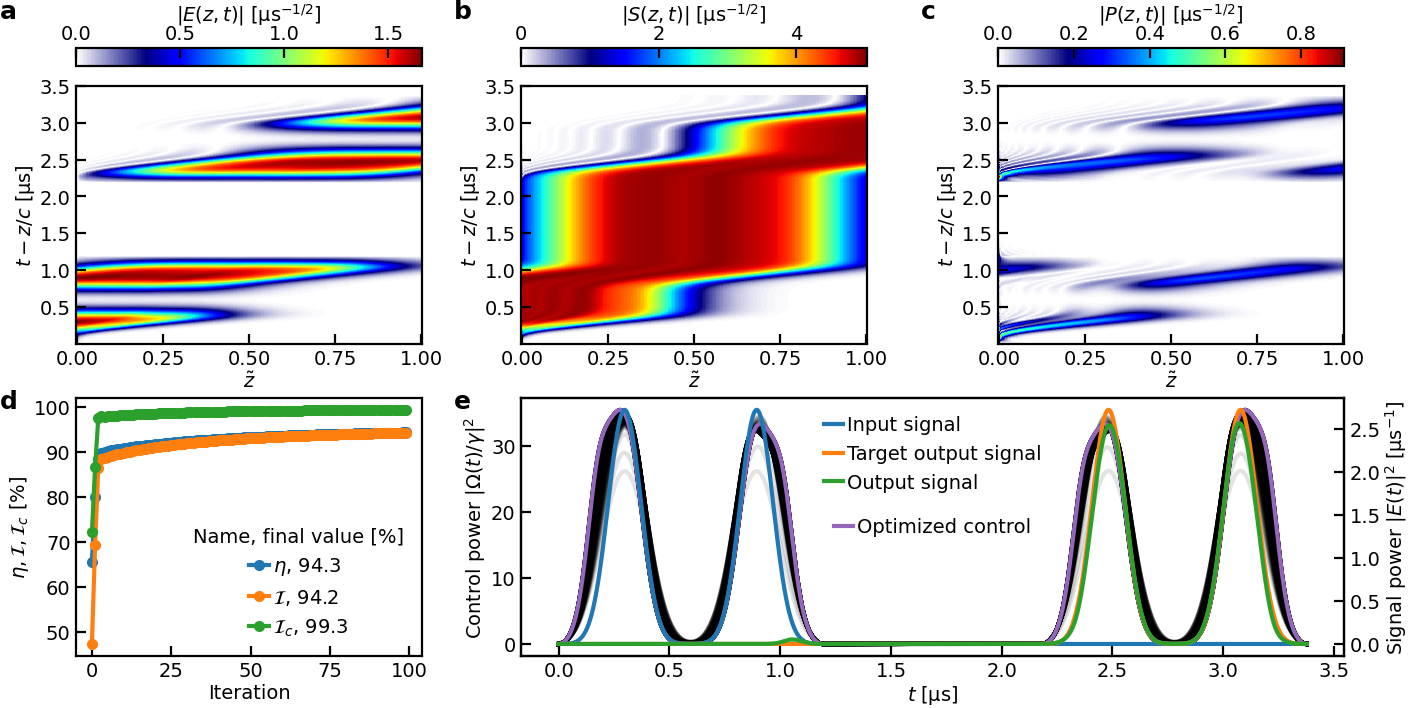}
    \caption{\textbf{Optimal waveforms and system dynamics corresponding to the two-Gaussian waveform for time-bin encoding in Fig.~4d of the main text.} See the caption of Sup. Fig.~\ref{fig:gauss_to_ion} for details.}
    \label{fig:timebin}
\end{figure}

Figures~\ref{fig:gauss_to_ion}, \ref{fig:gauss_to_10_gauss}, and \ref{fig:timebin} provides the system dynamics and the convergence for different examples of input and targeted output, as shown in Fig.~4 of the main text.

\clearpage
\section{Complete table of bandwidth adaptation data}

\begin{figure}[h]
    \centering
    \includegraphics[width=\textwidth]{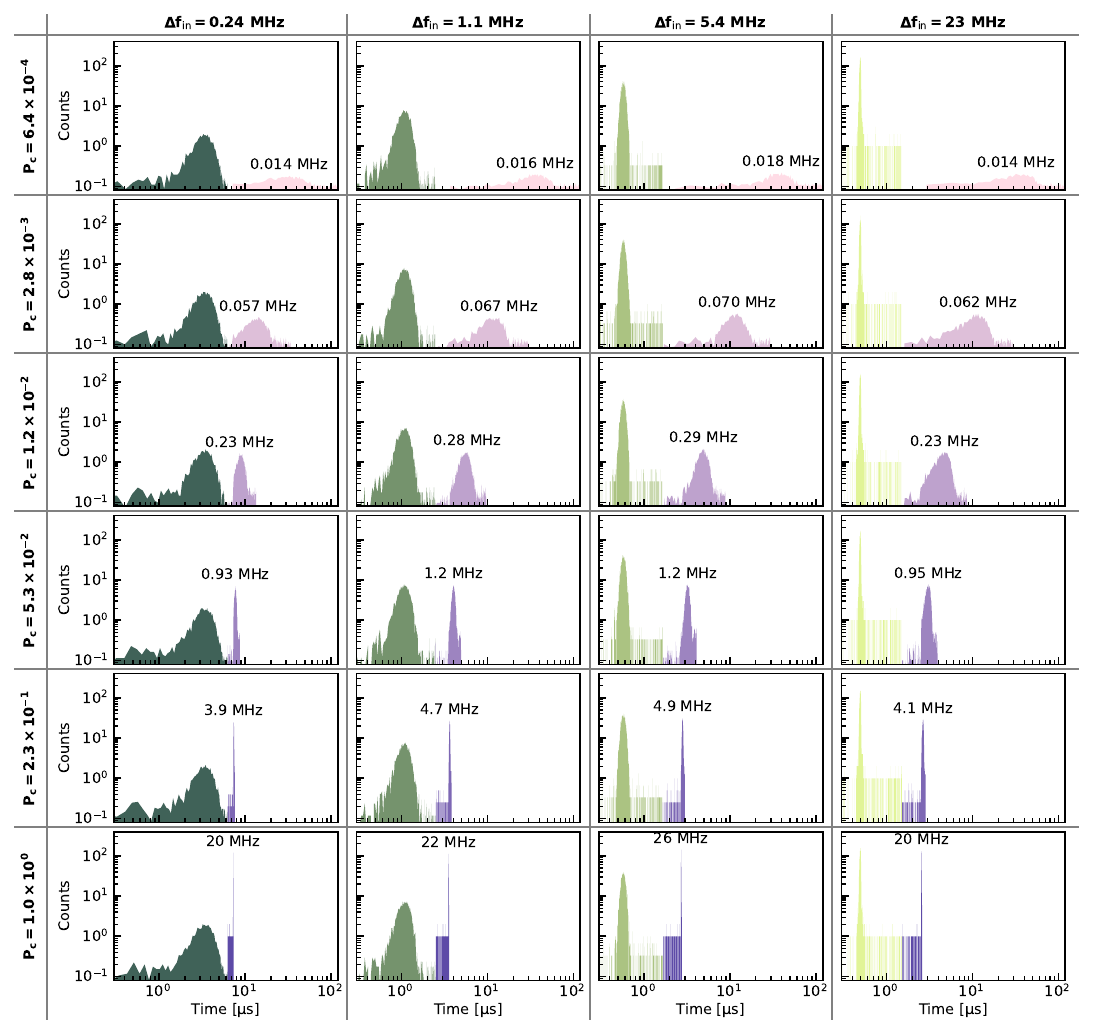}
    \caption{\textbf{Complete data used in Fig.~3 of the main text.} Conversion of all the input pulses $\abs{E_\mathrm{in}(t)}^2$, whose bandwidths are indicated in the column labels, to the output pulses $\abs{E_\mathrm{out}(t)}^2$ using the read control powers indicated in the row labels. The fitted bandwidth of the retrieved pulse $\Delta f_\mathrm{out}$ is indicated above the peak of each pulse.}
    \label{fig:table}
\end{figure}

For completeness, we show in Sup. Fig.~\ref{fig:table} the full set of bandwidth adaption plots used to produce Fig.~3 of the main text.

\clearpage
\bibliographystyle{naturemag}

\end{document}